\documentclass[aps,floatfix,preprintnumbers,letterpaper,ifmbe,twocolumn]{revtex4}
\usepackage{graphicx,framed}
\usepackage{epsfig}
\usepackage{amsfonts,amsmath,amsfonts,amscd}

\begin{document}

\title{Magnetofluid dynamics in curved spacetime}

\author{Chinmoy Bhattacharjee$^{\,1,2}$, Rupam Das$^{\,3}$,
S.M. Mahajan$^{\,1,2}$}

\affiliation{$^{1}$Institute for Fusion Studies, The University of
Texas at Austin, Austin,TX 78712 }

\affiliation{$^{2}$Department of Physics,The University of
Texas at Austin, Austin,TX 78712.}

\affiliation{$^{3}$Department of Physical and Applied Sciences, Madonna University, Livonia, MI 48150}

\begin{abstract}

A grand unified field $\mathcal{M}^{\mu\nu}$ is constructed from Maxwell's Field tensor and appropriately modified flow field, both non-minimally coupled to gravity, to analyze the dynamics of hot charged fluids in curved background space-time. With a suitable $3+1$ decomposition, this new formalism of the hot fluid is then applied to investigate the vortical dynamics of the system. Finally, the equilibrium state for plasma with non-linear coupling through Ricci scalar $R$ to gravity is investigated to derive a double Beltrami equation in curved space-time.

\end{abstract}

\maketitle
\section{Introduction}
By generalizing the standard minimum coupling prescription, $p_{\mu}\rightarrow  mU_{\mu}+qA_{\mu}$, invoked to incorporate the electromagnetic field in charged particle dynamics, an analogous theory to describe the dynamics of a hot charged fluid was developed in \cite{PhysRevLett.90.035001}. The ``hot-fluid" version of the prescription ($A_{\mu}$ is the electromagnetic four potential)
\begin{align}\label{coupling}
p_{\mu}\rightarrow  m\mathcal{G}U_{\mu}+qA_{\mu}= P_{\mu},
\end{align} 
combines the kinematic (the mass $m$, the four momentum $p_{\mu}$, and the four velocity $U_{\mu}$) and the statistical (the thermodynamic enthalpy $\mathcal{G}$) attributes of the fluid element. The centerpiece of the formalism was the construction of an antisymmetric, hybrid field tensor, 
\begin{align}\label{MTensor}
q M^{\mu\nu}= \nabla^{\mu} P^{\nu}-\nabla^{\nu} P^{\mu} = qF^{\mu\nu} + m S^{\mu\nu},
\end{align} 
that is a weighted sum of the electromagnetic field tensor $F^{\mu\nu}$, and the composite ( kinematic-statistical) fluid tensor  $S^{\mu\nu}= \nabla^{\mu} \mathcal{G}U_{\nu}-\nabla^{\nu} \mathcal{G}U_{\mu}$; the weighting factors are, respectively, the electric charge q, and the inertial ``charge" m. In analogy with electromagnetism, one may associate with $S^{\mu\nu}$, appropriately defined equivalents of the  electric and magnetic fields\cite{bekenstein1987helicity, PhysRevLett.90.035001}. 
The entire dynamics of the hot fluid is contained in the succinct equation (T is the temperature of the fluid)
\begin{align}\label{MTensorSR}
qU_{\mu}{M}^{\mu\nu}= \frac{mn\nabla^{\nu}\mathcal{G}-\nabla^{\nu}p}{n}=T\nabla^{\nu}\sigma, 
\end{align} 
where the right hand side is the thermodynamic force expressed in terms of the fluid entropy $\sigma$ using the standard thermodynamic relation between entropy with enthalpy ($p$ is the pressure). It was further shown that  the three-vector part of (\ref{MTensorSR}), after appropriate manipulation, is reducible to the conventional 3-D vortex dynamics except that the standard  fluid vorticity (and the conserved helicity) is replaced by the  hybrid magneto-fluid vorticity (magneto-fluid helicity). This is a far-reaching consequence because both the methodology and results of the highly investigated non-relativistic  vortex dynamics could, then, be transported to shed light on the much more complicated  hot relativistic fluid. 

In the light of the preceding discussion, Eq. (\ref{MTensorSR}) should be taken as a defining equation for the 4-D vortex dynamics.

A very fundamental result of the standard ``ideal" vortex dynamics is that it implies a topological invariant, the helicity (the hybrid-helicity), i.e.  such a dynamics forbids creation or destruction of vorticity; the vorticity could not be created from a state of no vorticity.  However, the constraint can be broken by including non-ideal behavior; in the standard 3D non relativistic system, the ideal behavior corresponds to the right hand side being a perfect gradient, $T\nabla^{\nu}\sigma=\nabla^{\nu}\bar{\sigma}$ which will require an equation of state of the type $\sigma=\sigma(T)$. It must be, however, emphasized that the thermodynamics of the fluid is its intrinsic property, and is not dictated by whether the equations of motion  can be cast in a ``canonical" vortex dynamics form.

The 4-D  vortex dynamics, though very similar to 3-D vortex dynamics, differed from it in one fundamental way; the special and general relativistic effects, through the distortion of space-time, could break the topological invariant even in ideal dynamics ($\sigma=\sigma(T))$. These effects  introduce  sources and sinks for  the relevant generalized helicity (ofter much more complicated than the 3-D fluid helicity)  so that  the creation and destruction of the generalized vorticity becomes possible in ideal dynamics. 

Thus the relativistic mechanics opens up a new vista and a new mode of analysis. We first try to construct the most general form of 4-D vortex dynamics (embodied in  Eq. (\ref{MTensorSR})) by manipulating the dynamics so that an appropriately generalized $\mathcal P^{\mu\nu}$ (and thus $\mathcal M^{\mu\nu})$ can be constructed to satisfy an equation of the type
\begin{align}\label{MTensorG}
qU_{\mu}\mathcal{M}^{\mu\nu}=  Q T\nabla^{\nu}\sigma, 
\end{align} 
where Q could be a function of spacetime geometry. As usual, the components $\mathcal M^{i j}=\epsilon^{i j k} X_k$ will define the new generalized vorticity (generalized magnetic field) $\bf{X}$.  And for homogeneous thermodynamics ($\nabla^{\nu}\sigma=0$), with or without an equation of state, the generalized helicity  $\mathcal H = <\vec{X}\cdot {\nabla\times}^{-1}\vec {X}>$  will be conserved. Here (${\nabla\times}^{-1}\vec {X}$) is the inverse curl of vortical field, $\vec{X}$.

Then the next step is to investigate relativistic mechanisms that, in combination with inhomogeneous thermodynamics, will create sources and sinks for $\mathcal{H}$, and $\bf{X}$. Finding the origin of seed vorticity (of the  magnetic field, for instance), which could be amplified in an ideal dynamo-like mechanism, is one of the most fascinating problems of theoretical astrophysics and cosmology, and several recent papers have advanced the effort by making the special relativistic  model of \cite{PhysRevLett.90.035001,PhysRevLett.105.095005,Mahajan_Yoshida} generally covariant  \cite{FAA,marklund2005general,Qadir:2014sca}, i.e, by  including gravity.

The general relativistic formulation in \cite{FAA}  was attempted within the framework of minimal coupling to gravity in the spherically symmetric and static space-time. Later the work is extended to rotating blackhole using the $\psi$-N (Pseudo-Newtonian formalism) framework \cite{Qadir:2014sca}.
In this  paper, we further generalize  the  recent work  by including non-minimal coupling between gravity and plasma in a general background space-time. 

Since our investigation facilitates a generalization to $f(R)$ gravity (not just Einstein's GR)\cite{Bertolami:2007gv,balakin2014nonminimal,Harko:2010zi,Harko:2014sja,Astashenok:2013vza}, we begin our analysis with an action functional for $f(R)$ gravity with perfect fluid and Maxwell's fields non-minimally coupled to gravity to derive the equation of motion for a new hybrid magnetofluid. Also, the vortical dynamics of the magnetofluid is explored by deriving generalized Faraday's law from dual tensor $\mathcal{M}^{*\mu\nu}$ and its  static spherically symmetric limits are explored  with their possible astrophysical applications. Finally, source free  plasma equilibrium states  in curved background space-time, are investigated.
In this paper, the calculation for generalized equation of motion of a new hybrid magnetofluid in curved background space-time is presented in the first section.  Next, the ADM formalism of electrodynamics\cite{thorne1982electrodynamics,thorne1986black,MTW,Wald} is applied to this new formulation of magnetofluids. Equations obtained after 3+1 decomposition are later cast into the vorticity evolution equation. Two limiting cases of the equation are shown later. Finally, the source free vorticity equation for the limiting cases with $f_m(R)=R$ is used to derive the equilibrium state of the system.

\section{Plasma Dynamics}\label{plasmadynamics}
The dynamics of an ideal plasma in curved background space-time can be investigated with the extremization of the following action functional in the convention $G = c = 1$,
\begin{widetext}
	\begin{equation*}
	S=S_{g}+S_{pfg}+S_M+S_{NM},
	\end{equation*}
	%\begin{widetext}
	\begin{equation}\label{action}
	S=\int d^4x \sqrt{-g}\left[\frac{1}{2} f_{g}(R)-(1+\lambda f_{m}(R))\rho(n,\tilde{\sigma})-\frac{1}{16\pi}(g^{\mu\alpha}g^{\nu\beta}+\mathcal{R}^{\mu\nu\alpha\beta})F_{\mu\nu}F_{\alpha\beta}\right],
	\end{equation}
\end{widetext}
where $S_g$, $S_{pfg}$, $S_{M}$, and $S_{NM}$ represent the corresponding action functionals for pure gravity, perfect fluid and maxwell's field minimally as well as non-minimally coupled to gravity. Since our analysis does not change in the context of modified gravity, we include,  for generality,  the functions of Ricci scalar $f_{g}(R)$, and $f_{m}(R)$ in the above action for pure gravity and coupling to matter respectively.
%The two terms in the action describe the minimal and non-minimal coupling of ideal plasma and electromagnetic field to gravity respectively.
Here $g_{\mu\nu}$, $g$, $F_{\mu\nu}$, $R$, and $R^{\mu\nu\alpha\beta}$ represent the metric tensor, the determinant of the metric tensor, the Maxwell field tensor, the Ricci scalar, and the Riemann tensor respectively.  The quantity $\rho(n,\tilde{\sigma})$, the energy density, is a function of number density, $n$, and entropy density, $\tilde{\sigma}$, of the plasma. Also, $\lambda$ is a phenomenological parameter that represents the coupling strength of the plasma to its background geometry; thus it can be treated as a coupling constant with a dimension of $length^2$ (see \cite{Astashenok:2013vza} for a detailed discussion). The non-minimal three parameter tensor $\mathcal{R}^{\mu\nu\alpha\beta}$ introduced in \cite{balakin2014nonminimal} has the form
\begin{equation*}
\mathcal{R}^{\mu\nu\alpha\beta}=q_1Rg^{\mu\nu\alpha\beta}+q_2\mathfrak{R}^{\mu\nu\alpha\beta}+q_3R^{\mu\nu\alpha\beta},
\end{equation*}
where  two auxiliary tensors  
\begin{equation*}
\mathfrak{R}^{\mu\nu\alpha\beta}=\frac{1}{2}(R^{\mu\alpha}g^{\nu\beta}-R^{\mu\beta}g^{\nu\alpha}+R^{\nu\beta}g^{\mu\alpha}-R^{\nu\alpha}g^{\mu\beta})
\end{equation*}
and
\begin{equation*}
g^{\mu\nu\alpha\beta}=\frac{1}{2}(g^{\mu\alpha}g^{\nu\beta}-g^{\nu\alpha}g^{\mu\beta}).
\end{equation*}
are introduced. The parameters $q_1$, $q_2$, and $q_3$ are used to describe non-minimal linear coupling of the electromagnetic tensor $F_{\mu\nu}$ to the curvature.  In general, these parameters are arbitrary and hence have to be chosen to satisfy some desired phenomenological or other constraints. For example, the Lagrangian with $q_1 = q_2 = 0$ and $q_3 = - \lambda_1$, has been used by Prasanna for phenomenological study of the non-minimal modifications of the electrodynamics \cite{Prasanna1971331,prasanna1,0264-9381-14-5-019}. Other suitable constraints for $q_1$, $q_2$, and $q_3$ have been chosen to befit the desired phenomenological or other results. Such a non-minimal coupling of electromagnetic field to gravity, with dimensional coupling constants  $q_1$, $q_2$, and $q_3$ as a natural choice of action, has been discussed by Balakin \cite{balakin2014nonminimal} and Horndeski \cite{horndeski}. Here, we keep all $q_1$, $q_2$, and $q_3$ without any constraint for the purpose of completeness and generality. One can choose a more generalized form of coupling of electromagnetic field to gravity through non-linear Riemann curvature, but we limit our analysis to the above action functional for simplicity.

Now, the extremization of the above action functional, i.e. varying it with respect to the metric $g_{\mu\nu}$, will result in the following modified Einstein's equation with the total stress-energy tensor $T^{\mu\nu}_{total}$ for the perfect fluid and Maxwell's field in curved space-time

\begin{widetext}
\begin{equation}\label{modified_eintsein}
F_{g}(R) R^{\mu\nu} - \frac{1}{2} f_{g}(R) g^{\mu\nu}-(\nabla^{\mu}\nabla^{\nu}-g^{\mu\nu}\square)F_{g}(R)= 8\pi T^{\mu\nu}_{total},
\end{equation}
\end{widetext}
where $F_{g}=f_{g}'(R)$ (differentiation with respect to $R$) and  $ T^{\mu\nu}_{total}$ is the total stress-energy tensor computed from $ T^{\mu\nu}=(2/\sqrt{-g})\left(\delta S/\delta g_{\mu\nu}\right)$. 
It is straightforward to show that the Einstein's equation for GR is recovered by setting $f_{g}(R)=R$ and $F_{g}(R)=1$ and the divergence of the modified Einstein's equation (\ref{modified_eintsein}) produces the equation of motion of the plasma in curved background space-time since the vanishing divergence of the left hand side will not contribute to the equation of motion for the plasma.\\

The stress energy tensor for perfect fluid and Maxwell's field can be computed from $ T^{\mu\nu}=(2/\sqrt{-g})\left(\delta S/\delta g_{\mu\nu}\right)$ with the corresponding action functional. Varying the action functional $S_{pfg}$, we obtain the following stress tensor for perfect fluid coupled to gravity \cite{Bertolami:2011rb,Bertolami:2008im,Bertolami:2008zh,Bertolami:2007gv}
\begin{eqnarray}\label{fluidstress}
T^{\mu\nu}_{pfg}=(1+\lambda f_{m}(R)) T^{\mu\nu}_{pf} + 2\lambda \rho F_{m} (R)R^{\mu\nu} \notag \\   -2\lambda (\nabla^{\mu}\nabla^{\nu}-g^{\mu\nu}\square)\rho F_{m}(R), 
\end{eqnarray}
where  $T^{\mu\nu}_{pf}$ is the stress-energy tensor for non-coupled perfect fluid (See Appendix \ref{pfstresstensor} for details):
$ T^{\mu\nu}_{pf}=(p+\rho)U^{\mu}U^{\nu}+pg^{\mu\nu}$ with $U^{\mu}=dx^{\mu}/{d\tau}$ being the four velocity of the plasma particles, and $F_{m}(R)=f_{m}'(R)$. The quantity $p+\rho=h$ is known to be the enthalpy density of the plasma, which can be expressed by introducing an auxiliary function $\mathcal{G}=h/mn$ with $m$ and $n$ being the mass and number density respectively. \\

Using the identity $\nabla_{\mu}(\nabla^{\mu}\nabla^{\nu}-g^{\mu\nu}\square)\rho=R^{\mu\nu}\nabla_{\mu}\rho$, the expression for the divergence of the stress tensor is \cite{Bertolami:2007gv,Harko:2010zi,Harko:2014sja,Harko:2014gwa,Bertolami:2008zh,Bertolami:2008im,Bertolami:2011rb},
\begin{equation}\label{fluiddiv}
\nabla_{\mu}T^{\mu\nu}_{pfg}=(1+\lambda f_{m}(R))\nabla_{\mu}T^{\mu\nu}_{pf}+\lambda F(T^{\mu\nu}_{pf}+g^{\mu\nu}\rho)\nabla_{\mu}R.
\end{equation}

On the other hand,  the variation of the action functional $S_M$ gives the usual electromagnetic stress tensor
\begin{equation}\label{emstress}
T^{\mu\nu}_M=\frac{1}{4\pi}\left(F^{\nu\beta}F^{\mu}\ _{\beta}-\frac{1}{4}F_{\mu\nu}F^{\mu\nu}\right),
\end{equation}
and, with the Bianchi identity, the divergence of $T^{\mu\nu}_{M}$ takes the following form
\begin{equation}\label{divem}
4\pi \nabla_{\mu}T^{\mu\nu}_M=-F^{\nu} \ _{\beta}\nabla_{\alpha}F^{\beta\alpha}.
\end{equation}
\\
Next, owing to the linear nature of non-minimal action, we can assume the total non-minimal stress-energy tensor to be \cite{balakin2014nonminimal}
\begin{equation*}
T_{NM}^{\mu\nu}=q_1T_{1}^{\mu\nu}+q_2 T_{2}^{\mu\nu}+q_3 T_{3}^{\mu\nu}.
\end{equation*} 
Varying the action for non-minimal coupling, we obtain the following set of three stress energy tensors
\begin{widetext}
	\begin{align}\label{nonminstress1}
	T_1^{\mu\nu}=\frac{1}{4\pi}\Bigl[-\frac{1}{2}(\nabla^{\mu}\nabla^{\nu}-g^{\mu\nu}\square)F_{\alpha\beta}F^{\alpha\beta}+RF^{\mu \beta}F^{\nu} \ _{\beta}\Bigr],\end{align}
	%\begin{widetext}
	\begin{align}\label{nonminstress2}
	T_2^{\mu\nu}= \ &\frac{1}{4\pi} \ g^{\mu\alpha}g^{\nu\beta}\Bigl[\frac{1}{2}g_{\alpha\beta}\left(\nabla_{\gamma}\nabla_{\theta}(F^{\gamma \sigma}F^{\theta}\ _{\sigma})-R^{\gamma\sigma}F_{\gamma \theta}F^{\theta}\  _{\sigma}\right)+F^{\gamma \sigma}(R_{\gamma\beta}F_{\alpha \sigma}+R_{\gamma\alpha}F_{\beta \sigma}) \notag \\
	& +\frac{1}{2}\square (F_{\beta \sigma}F_{\alpha}\ ^{\sigma} )-\frac{1}{2}\nabla_{\gamma }\left[\nabla_{\alpha}(F_{\beta \sigma}F^{\gamma \sigma})+\nabla_{\beta}(F_{\alpha \sigma}F^{\gamma \sigma})\right]+R^{\gamma\sigma}F_{\gamma\alpha}F_{\sigma\beta}\Bigr],
	\end{align}
	\begin{align}\label{nonminstress3}
	T^{\mu\nu}_{3}= \ & \frac{1}{4\pi} \Bigl[\ g^{\mu\alpha}g^{\nu\beta}\Bigr(-\frac{1}{4}g_{\alpha\beta}R^{\gamma\theta\sigma\rho}F_{\gamma\theta}F_{\sigma\rho}+\frac{3}{4}F^{\sigma\rho}(F_{\alpha} \ ^{\theta}R_{\beta\theta\sigma\rho}+F_{\beta} \ ^{\theta}R_{\alpha\theta\sigma\rho})\notag \\
	&+\frac{1}{2}\nabla_{\gamma}\nabla_{\theta}[F_{\alpha}\ ^{\gamma}F_{\beta}\ ^{\theta}+F_{\beta}\ ^{\gamma}F_{\alpha}\ ^{\theta}]\Bigr)\Bigr].
	\end{align}
\end{widetext}

The calculation of the divergences of the above three stress tensors requires electrodynamic equations corresponding to the total action functional, i.e. adding an interaction or source term in the action. Therefore, we take the variation of the action functional with the source term with respect to the field variable $A^{\mu}$ and obtain \cite{balakin2014nonminimal}
\begin{align}\label{maxwelleq}
\nabla_{\alpha}H^{\alpha\beta}=-4\pi nq U^{\beta},
\end{align}
where
\begin{align}
\label{grandtensor}
H^{\alpha\beta}=F^{\alpha\beta}+\mathcal{R}^{\mu\nu\alpha\beta}F_{\mu\nu}
\end{align}
may be considered as a generalized Faraday tensor in curved background space-time and $nqU^{\beta}$ is the Lorentz four current associated with the charged fluid. Equations(\ref{maxwelleq}) and (\ref{grandtensor}) can be regarded as the constitutive relations of the unified system to preserve the Maxwell's equations in curved background space-time. \\

Invoking the fact the  total stress energy tensor should be divergence free, i.e.
\begin{align}
\label{mastereq}
\nabla_{\mu}\Bigl[T^{\mu\nu}_{pfg}+T^{\mu\nu}_{M}+T^{\mu\nu}_{NM}\Bigr]=0,
\end{align}
with $T^{\mu\nu}_{NM}=q_1T_1^{\mu\nu}+q_2T_2^{\mu\nu}+q_2T_2^{\mu\nu}$, and using equations (\ref{divem}), (\ref{grandtensor}) and (\ref{mastereq}), we obtain the  divergences of the three constituents of the non-minimal stress energy tensor $T_{NM}^{\mu\nu}$:
\begin{align}\label{nonminstressdiv1}
4\pi\nabla_{\mu}T^{\mu\nu}_1=-F^{\nu} \ _{\beta}\nabla_{\alpha}(RF^{\alpha\beta}),
\end{align}
\begin{align}\label{nonminstressdiv2}
4\pi\nabla_{\mu}T^{\mu\nu}_2=-F^{\nu} \ _{\beta}\nabla_{\mu}(R^{\mu\gamma}F_{\gamma}\ ^{\beta}+R^{\gamma\beta}F^{\mu} \  _{\gamma}),
\end{align}
\begin{align}\label{nonminstressdiv3}
4\pi\nabla_{\mu}T^{\mu\nu}_3=-F^{\nu} \ _{\beta}\nabla_{\mu}(R^{\mu\beta\gamma\theta}F_{\gamma\theta}).
\end{align}
Substituting these expressions for divergence in (\ref{mastereq}), we obtain
\begin{widetext} 
	\begin{align}\label{eom1}
	(1+\lambda f_{m}(R))\nabla_{\mu}T^{\mu\nu}_{pf}=\left[qnF^{\nu} \ _{\beta}U^{\beta}-\lambda F_m(R)(T^{\mu\nu}_{pf}+g^{\mu\nu}\rho)\nabla_{\mu}R\right].
	\end{align}
\end{widetext}
It is interesting to note that the coupling between gravity and electromagnetic field is now explicitly manifest in Eq.(\ref{grandtensor}) through generalized Faraday tensor, and is implicit in the equation of motion,  (\ref{eom1}), for plasma in curved background space-time through the current since, unlike standard electromagnetic field, $U^\beta$ is governed by a hybrid field $H^{\alpha\beta}$.

Until now, we mainly followed the standard approach to derive the covariant equation of motion for the plasma in curved background space-time from the action functional (\ref{action}). The main result is equation (\ref{eom1}) that captures the non-minimally gravity-coupled plasma dynamics. To advance in our program of unifying the electromagnetic field with an appropriately weighted flow field, we next derive a generalized expression for the corresponding unified magnetofluid.

\subsection{Magnetofluid Unification} \label{unification}
Following the prescription presented in \cite{PhysRevLett.90.035001}, we substitute the expression for the stress tensor $T^{\mu\nu}_{p f}=(p+\rho)U^{\mu}U^{\nu}+pg^{\mu\nu}$ for perfect fluid in equation (\ref{eom1}) and invoke the continuity equation $\nabla_{\mu}(nU^{\mu})=0$ to obtain
\begin{widetext}
	\begin{eqnarray}\label{eom2}
	 (1+\lambda f_{m}(R)-\lambda RF_{m}(R))mnU^{\mu}\nabla_{\mu}(\mathcal{G}U^{\nu})+(1+\lambda f_{m}(R)-\lambda RF_{m}(R))\nabla^{\nu}p \notag \\
 +\lambda mnF_{m}(R)U^{\mu}\nabla_{\mu}(R\mathcal{G}U^{\nu})  +\lambda F_{m}(R)\nabla^{\nu}(pR)=qn F^{\nu\beta}U_{\beta}-\lambda F_{m}(R) \rho\nabla^{\nu}R.
        \end{eqnarray}
\end{widetext}
in terms of the standard perfect fluid flow tensor $S^{\mu\nu}=\nabla^{\mu}(\mathcal{G} U^{\nu})-\nabla^{\nu}(\mathcal{G}U^{\mu})$, and a new curvature-coupled weighted antisymmetric flow field tensors $K^{\mu\nu}=\nabla^{\mu}(R\mathcal{G}U^{\nu})-\nabla^{\nu}(R\mathcal{G}U^{\mu})$, we can manipulate equation (\ref{eom2}) to obtain ($\mathcal{G}=h/mn$)  
\begin{equation}\label{eom3}
(1+\lambda f_{m}(R))\left[\frac{\nabla^{\nu}p}{n}-m \nabla^{\nu}\mathcal{G}\right]=qU_{\mu}\mathcal{M}^{\nu\mu},
\end{equation}
where the new grand vorticity tensor has the ``canonical" form (\cite{PhysRevLett.90.035001})
\begin{equation}\label{unifiedtensordef}
\mathcal{M}^{\nu\mu}=F^{\nu\mu}+\frac{m}{q}D^{\nu\mu}
\end{equation} with
\begin{equation}\label{unifiedtensordef}
D^{\nu\mu}=(1+\lambda f_{m}(R)-\lambda RF_m)S^{\nu\mu}+\frac{m}{q}\lambda F_mK^{\nu\mu}.
\end{equation}
The new fluid tensor $D^{\mu\nu}$ displays, explicitly, the coupling of flow field to gravity. Using the thermodynamic identity 
\begin{equation}\label{identity}
\nabla^{\nu}\sigma=\frac{mn\nabla^{\nu}\mathcal{G}-\nabla^{\nu}p}{nT},
\end{equation}
we cast equation (\ref{eom3}), governing the dynamics of a hot fluid system in curved background space-time, into the  ``canonical" 4-D vortex form 
\begin{equation}\label{finaleom}
qU_{\mu}\mathcal{M}^{\mu\nu}=(1+\lambda f_{m}(R))T\nabla^{\nu}\sigma.
\end{equation}
Here it must be noted that invoked thermodynamic relation is contingent  upon an appropriately well-defined local concept of temperature in curved space-time. The above thermodynamic identity (\ref{identity}) can be derived from the first law of thermodynamics expressed in the form of exact differential $dH=m d\mathcal{G}=Td\sigma + Vdp=Td\sigma + dp/n $, with $H$ and $V$ being the enthalpy and the volume of the fluid element respectively, which in turn can be cast into the following form $m\nabla^{\mu}\mathcal{G} =T\nabla^{\mu}\sigma+(1/n)\nabla^{\mu} p$ for a fluid element moving along the worldline with four velocity $U^{\mu}$ \cite{MTW}.

Notice that,  when $\lambda = 0$, $\mathcal{M}^{\mu\nu}$ reduces to its minimally-coupled  counterpart-the tensor $M^{\mu\nu}$ defined in \cite{PhysRevLett.90.035001,FAA}. 

%It is also obvious from the equations (\ref{finaleom}) and (\ref{entropydef}) that the magnetofluid field reaches a degenerate (force free) state for fluids with its equation of state obeying $\nabla^{\nu}{\sigma}=0$, for example, the homentropic fluids with curvature coupling $\lambda f_{m}(R) \neq -1$.  %This is not surprising since the magnetofluid field $\mathcal{M}^{\mu\nu}$ itself implicitly depends on curvature. Therefore, the more explicit condition for degenerate states is $\nabla^{\nu}p = mn\nabla^{\nu}\mathcal{G}$.\\ 

Equation (\ref{finaleom}) is the main result of this formalism; we have just shown that a charged relativistic fluid, coupled non-minimally to gravity, obeys a 4-D vortex dynamics like its gravity free, and minimally coupled to gravity,  counterparts. The new grand vorticity tensor subsumes earlier limiting cases in  transparent manner.

We would now apply the above covariant formulation to spell out and investigate the more advanced  vortical structures contained in this system. To do calculations in terms of familiar quantities, we will  begin with a  3+1 decomposition.

\section{3+1 dynamics of gravito-magnetofluid }\label{magnetofluiddynamics}

The 3+1 decomposition of the 4-D  vortex dynamics will help us, inter alia, to  find: 1) generalized electric and magnetic field from $\mathcal{M}^{\mu\nu}$, and 2) the energy and the continuity equation rewritten in terms of generalized electric and magnetic fields. \\

The approach chosen for the 3+1 splitting selects a family of foliated fiducial 3-dimensional hypersurfaces (slices of simultaneity) $\Sigma_{t}$ labelled by a parameter $t = constant$ in terms of a time function on the manifold. Furthermore, we let $t^{\mu}$ be a timeline vector whose integral curves intersect each leaf $\Sigma_{t}$ of the foliation precisely once and which is normalized such that $t^{\mu}\nabla_{\mu}t = 1$. This $t^{\mu}$ is the `evolution vector field' along whose orbits different points on all  $\Sigma_{t} \equiv \Sigma$ can be identified. This allows us to write all space-time fields in terms of $t$-dependent components defined on the spatial manifold $\Sigma_{t}$. Lie derivatives of space-time field along $t^{\mu}$ are identified with `time-derivatives' of the spatial fields since Lie derivatives reduce to partial time derivative for an adapted coordinate system $t^{\mu}=(1,0,0,0)$.

Moreover, since we are using the Lorentzian signature, the vector field $t^{\mu}$ is required to be future directed. Let us decompose $t^{\mu}$ into normal and tangential parts with respect to $\Sigma_{t}$ by defining the lapse function $\alpha$ and the shift vector $\beta^{\mu}$ as $t^{\mu}=\alpha n^{\mu}+\beta^{\mu}$ with $\beta^{\mu}n_{\mu} = 0$, where  $n^{\mu}$ is the future directed unit normal vector field to the hypersurfaces $\Sigma_{t}$. More precisely, the natural timelike convector $n_{\mu}=(-\alpha,0,0,0)=-\alpha\nabla_{\mu}t$  is defined to obtain $n^{\mu}=({1}/{\alpha},-\beta^{\mu}/\alpha)$ which satisfy the normalization condition $n^{\mu}n_{\mu}=-1$. Then, the space-time metric $g_{\mu\nu}$ induces a spatial metric $\gamma_{\mu\nu}$ by the formula $\gamma_{\mu\nu}=g_{\mu\nu}+n_{\mu}n_{\nu}$. Finally, the 3+1 decomposition is usually carried out with the projection operator $\gamma^{\mu}\ _{\nu}=\delta^{\mu} \ _ {\nu}+n^{\mu}n_{\nu}$, which satisfies the condition $n^{\mu}\gamma_{\mu\nu}=0$. Also, the acceleration is defined as $a_{\mu}=n^{\nu}\nabla_{\nu}n_{\mu}$. \\

Now with the above foliation of space-time, the space-time metric  takes the following canonical form \cite{MTW}
\begin{equation}\label{canonicalmetric}
ds^2=-\alpha^2dt^2+\gamma_{ij}(dx^i+\beta^i dt)(dx^j+\beta^j dt),
\end{equation}
and it immediately follows that, with respect to an Eulerian observer, the Lorentz factor turns out to be
\begin{equation}\label{lorentzfactor}
\Gamma=\left[\alpha^2-\gamma_{ij}(\beta^{i}\beta^{j}+2\beta^{i}v^{j}+v^{i}v^{j})\right]^{-1/2},
\end{equation}
satisfying $d\tau = dt/\Gamma$, where $v^i$ is the $i$th component of fluid velocity $\vec{v}=d\vec{x}/dt$.  Then the decomposition for the four velocity is \cite{FAA}
\begin{equation}\label{fourvelocity}
U^{\mu}=\alpha \Gamma n^{\mu}+\Gamma\gamma^{\mu} \ _{\nu}v^{\nu},
\end{equation}
with $n_{\mu}U^{\mu}=-\alpha \Gamma$.

Now, since our unified anti-symmetric field tensor $\mathcal{M}^{\mu\nu}$ is constructed from the anti-symmetric tensors $F^{\mu\nu}$ and $D^{\mu\nu}$, we apply the ADM formalism of electrodynamics presented in \cite{thorne1982electrodynamics,thorne1986black,MTW,Wald} to define the generalized electric and magnetic field respectively as 
\begin{alignat}{2}\label{generalemdef}
\xi^{\mu}=n_{\nu}\mathcal{M}^{\mu\nu}\ ;\  &\qquad\text{}\qquad X^{\mu}=\frac{1}{2}n_{\rho}\epsilon^{\rho\mu\sigma\tau}\mathcal{M}_{\sigma\tau},
\end{alignat}
and thus express the unified field tensor
\begin{equation}\label{generalfieldtensordef}
\mathcal{M}^{\mu\nu}=n^{\mu}\xi^{\nu} - n^{\nu}\xi^{\mu}-\epsilon^{\mu\nu\rho\sigma}X_{\rho}n_{\sigma}.
\end{equation}
We remind the reader that the generalized magnetic field and the generalized vorticity are essentially synonymous. Using the definition of the unified field tensor $\mathcal{M}^{\mu\nu}$,  the expressions of 3D generalized electric and magnetic field turn out to be
\begin{widetext}
	\begin{align}\label{generalefield}
	\vec{\xi}= \ & \vec{E}-\frac{m}{q}(1+\lambda f_{m}(R)-\lambda RF_{m}(R))\vec{\nabla}(\alpha \mathcal{G}\Gamma)-\frac{m}{q}\lambda F_{m}(R)\vec{\nabla}(\alpha \mathcal{G}R\Gamma)-\frac{m}{q}(1+\lambda f_{m}(R))\left[2\underline{\underline{{\sigma}}}\cdot(\mathcal{G}\Gamma\vec{v})+\frac{2}{3} \theta\mathcal{G}\Gamma\vec{v}\right]\notag \\
	& -\frac{m}{q\alpha}(1+\lambda f_{m}(R)-\lambda RF_{m}(R))\left(\mathcal{L}_t(\mathcal{G}\Gamma\vec{v})-\mathcal{L}_{\vec{\beta}}(\mathcal{G}\Gamma\vec{v})\right)-\frac{m}{q\alpha}\lambda F_{m}(R)\left(\mathcal{L}_t(\mathcal{G}R \Gamma\vec{v})-\mathcal{L}_{\vec{\beta}}(\mathcal{G}R \Gamma\vec{v})\right);
	\end{align}
	\begin{align}\label{generalbfield}
	\vec{X}=\vec{B}+\frac{m}{q}(1+\lambda f_{m}(R)-\lambda RF_{m}(R))\ \vec{\nabla}\times(\mathcal{G}\Gamma\vec{v})+\lambda F_{m}(R) \frac{m}{q}\vec{\nabla}\times(R\mathcal{G}\Gamma\vec{v}),
	\end{align}
\end{widetext}
where $\underline{\underline{{\sigma}}} = \sigma_{\mu}^{\nu}$ and $\theta$ are, respectively,   the shear and expansion of the congruence, defined as $\sigma_{\alpha\beta} = \gamma_{\alpha}^{\mu}\gamma_{\beta}^{\nu}\nabla_{(\mu} n_{\nu)}- \frac{1}{3}\theta \gamma_{\mu\nu}$ and $\theta =\nabla_{\mu}n^{\mu} $. We have also used the relation $\nabla_{\mu} n_{\nu} = -a_{\nu}n_{\mu} + \sigma_{\alpha\beta} + \frac{1}{3}\theta \gamma_{\mu\nu}$ to derive (\ref{generalefield}).

Finally the $\gamma^{\beta}\ _{\mu}$ projection of the unified field equation of motion (\ref{finaleom}) gives us the momentum evolution equation
\begin{equation}\label{generalmomeq}
\alpha q \Gamma \vec{\xi}+q\Gamma(\vec{v}\times\vec{X})=-(1+\lambda f_{m}(R))T\vec{\nabla}\sigma
\end{equation}
whereas the $n_{\mu}$ projection gives the equation of energy conservation
\begin{equation}\label{generalenergyeq}
\alpha q\Gamma \vec{v}\cdot\vec{\xi}=T (1+\lambda f_{m}(R))(\mathcal{L}_t\sigma-\vec{\beta}\cdot\vec{\nabla}\sigma).
\end{equation}

\subsection{Vortical Dynamics}\label{vorticialdynamics}
Understanding the  full extent of this formalism is, perhaps, a very long term project. We can, however, begin to appreciate its rich content by exploring some aspects of the 
new vortical dynamics.   Sources responsible for magnetic field generation, in particular, the sources that are gravity driven, can be derived by deriving the generalized vorticity evolution equation (which is really the generalized Faraday's law) by manipulating Eq. (\ref{generalmomeq}). 

Since $\mathcal{M}^{\mu\nu}$ is an anti-symmetric tensor, the divergence of its dual is zero, i.e. $\nabla_{\mu}\mathcal{M}^{*\mu\nu}=0$. Taking  the $\gamma^{\beta}\ _{\mu}$  projection of the preceding identity, we derive

	\begin{align}\label{faradaygeneral}
	\mathcal{L}_t\vec{X}=\mathcal{L}_{\vec{\beta}}\vec{X}-\vec{\nabla}\times(\alpha\vec{\xi})-\alpha \theta \vec{X},
	\end{align}\\
where $\mathcal{L}$  denotes Lie derivatives with $\mathcal{L}_t = \partial_t$ along $t^{\mu}$ and  $\mathcal{L}_{\vec{\beta}}{\vec{X}}= [\vec{\beta},\vec{X}]$.
	
It should be noted that even in the absence of non-minimal coupling to gravity ($\lambda=0$), minimal coupling to gravity, is always present. 
Equation (\ref{faradaygeneral}), in conjunction with equation  (\ref{generalmomeq}), gives us the vorticity evolution equation of the system
\begin{widetext}
	\begin{align}\label{vorticity}
	\mathcal{L}_t\vec{X}-\vec{\nabla} \times (\vec{v}\times \vec{X})-\mathcal{L}_{\vec{\beta}}\vec{X}+\alpha \theta \vec{X}=\vec{\nabla}\times\left(\frac{T}{q\Gamma}(1+\lambda f_{m}(R))\vec{\nabla}\sigma\right).
	\end{align}
\end{widetext}

All terms on the left hand side  operate on the vorticity three-vector $\vec{X}$ while the right hand side  provides, just as in the conventional picture, possible sources for vorticity generation.  
The left hand side, however, has  lot more structure than the conventional 3-D vortex dynamics; the first two terms are the standard Helmholtz like, 
while $\alpha \theta \vec{X}$ and $\mathcal{L}_{\vec{\beta}}\vec{X}$, are nontrivial gravity modifications. Thus the gravity coupling does, fundamentally, modify the projected 3D
vortex dynamics, in spite of the fact, that the 4D vortex equations had exactly the same form.

Till now, our analysis has been very general with the assumption that the space-time structure satisfies the modified Einstein's equation (\ref{modified_eintsein}) and it permits the 3+1 foliation adopted above. Further investigation is better done after specifying the precise structure of space-time.  Without the knowledge of the structure of space-time, it may not be possible, even, to specify conditions under which the helicity, a topological invariant of the system, is conserved \cite{Pino:2009vj,PhysRevLett.90.035001,PhysRevLett.105.095005}. Since it is beyond the scope of our current endeavor to find the solutions of the modified Einstein's equation (\ref{modified_eintsein}) to explore the vortical dynamics, we instead briefly discuss the vortical dynamics in the context of a couple of well known space-time solutions to the original Einstein's equation.

First, for a minimal coupling ($\lambda =0$) and a spherically symmetric and static space-time like Schwarzchild solution, the above vortical evolution equation (\ref{vorticity}) reduces to the one presented in \cite{FAA}, i.e. $\mathcal{L}_t\vec{X}-\vec{\nabla} \times (\vec{v}\times \vec{X})=\vec{\nabla}\times\left((T/q\Gamma)\vec{\nabla}\sigma\right)$. Since the spherically symmetric and static space-time can be foliated without the shift function $\vec{\beta}$, and the foliation obeys the time translation symmetry leading to a vanishing extrinsic curvature, the two new terms on the left hand side disappear. Thus the structure is precisely like  the 3-D vortex dynamics. The simplified vortical evolution equation can be used to approximately compute the weak field seed generation in  the hot fluid system in the accretion disc of the Schwarzchild black hole \cite{FAA}. 

Second, for a non-minimal coupling and a spherically symmetric and static space-time, the above vortical evolution equation (\ref{vorticity}) again reduces to the 3-D like vortex dynamics , i.e. $\mathcal{L}_t\vec{X}-\vec{\nabla} \times (\vec{v}\times \vec{X})=\vec{\nabla}\times\left((T/q\Gamma)(1+\lambda f_{m}(R))\vec{\nabla}\sigma\right)$. Again, spherical symmetry and non-rotating space-time demand that the two terms on the left hand side corresponding to the shift function and the expansion factor disappear and thus render the vortical evolution equation to be applied for computing seed generation in massive astrophysical objects, especially with $f_{m}(R)=R$. The equilibrium plasma state in  this space-time  will be  discussed  a little later.

Finally, for $f_{m}(R)=R$ and a stationary and axially symmetric space-time such as Kerr black hole, the entire vortical evolution equation (\ref{vorticity}) with appropriate modifications can be used to explore a number of astrophysical applications including gamma ray bursts and seed generation, which will be explored in the near future. Therefore, the study of plasma dynamics in curved background space-time in light of the newly constructed grand unified field tensor $\mathcal{M}^{\mu\nu}$ provides several new useful insights that can be used to explore some astrophysical phenomena.

\subsection{Equilibrium state}
For $f_m(R)=R$, we can explore the source free vorticity evolution equation with non-zero $\lambda$, $\mathcal{L}_t\vec{X}-\vec{\nabla} \times (\vec{v}\times \vec{X})=0$. The trivial solution for this equation is
$\vec{X}=\vec{B}+(mc/q)\vec{\nabla}\times((1+\lambda R)\mathcal{G}\Gamma \vec{v})=0$. Using the GR modified Ampere's law $(\vec{\nabla}\times\alpha \vec{B})=(4\pi/c) (\alpha
q n\Gamma\vec{v})$, we obtain the equation for equilibrium state
\begin{align}\label{equil_state}
\vec{\nabla}\times\vec{\nabla}\times \vec{B}=\ & -\frac{1}{\mathcal{Q}\lambda^2_L}\vec{B} -\vec{\nabla}\ln (\mathcal{Q})\times(\vec{a}\times\vec{B})\notag \\
 & -\vec{\nabla}\times(\vec{a}\times\vec{B})-\vec{\nabla}\ln (\mathcal{Q})\times(\vec{\nabla}\times \vec{B}),
\end{align}
where $\mathcal{Q}=\mathcal{G}(1+\lambda R)$, skin depth $\lambda^2_L=c^2/\omega_p^2$, plasma frequency $\omega_p^2=4\pi n q^2/m$ and acceleration $\vec{a}=\vec{\nabla}\ln\alpha$. Without gravity terms, we see that equation (\ref{equil_state}) becomes the familiar london equation, i.e. $\vec{\nabla}\times\vec{\nabla}\times \vec{B}= -(1/\lambda^2_L)\vec{B}$. However, with gravity entering the system with minimal and non-minimal coupling, we observe that the equilibrium state is more restrictive than the corresponding classical system. Terms with $\alpha$ and $\lambda R$ on the right hand side of the above equation (\ref{equil_state}) show contributions from minimal and non-minimal coupling respectively. The acceleration $\vec{a}$ in the equation stands for the gravitational force felt by plasma as it goes from one to the next hypersurface. It is interesting to notice the explicit appearance of the interaction between Maxwell's field and gravity through the equilibrium state that was previously hidden in the definition of four current $I^{\beta}$ mentioned in Section \ref{plasmadynamics}. 

It must be emphasized that equation (\ref{equil_state}) is a grand generalization of the London equation (the canonical vorticity being zero) and, thus, acquires much more structure  due to gravity- a kind of generalized superconductivity in which the grand vorticity is expelled from the interior. Such states, belonging to the well-known category of relaxed states obtained by satisfying the constraint $\vec{\nabla}\cdot\vec{B}=0$ and by imposing appropriate boundary conditions that are dependent on specific geometry of the system, may be used to model the equilibria of  plasmas coupled to a strongly gravitating sources. We could, for example, seek localized solutions ($\vec{B}=0$ as $r\rightarrow \infty$) for a black hole accretion disk, an example which is discussed briefly in the context of vorticity generation. A complete analysis of this equation will  depend on many aspects:  spacetime geometry, temperature profile and most importantly a profile for plasma frequency as number density in the accretion disk is a complicated function of distance which also varies in different regions of the disk \cite{Shakura}. Our future work will explore a complete analysis for the equilibrium state that will require substantial numerical analysis.

\section{Conclusion}
As a next step to the the unified theory of electromagnetic fields and flow fields (\cite{PhysRevLett.90.035001, FAA}), we have constructed a formalism describing the dynamics of hot charged relativistic fluids, non-minimally coupled  to gravity. It is shown that, even with non-minimal coupling, the dynamics obeys the 4-D vortical structure, first exposed in equation (\ref{finaleom})\cite{PhysRevLett.90.035001}. The new vorticity tensor $\mathcal{M}^{\mu\nu}$ represents a grand synthesis of fluid, electromagnetic and modified gravity fields with  the non-minimal gravity coupling appearing, explicitly, in its definition. The current formalism, when expressed in  3+1 decomposition incorporates  shear, and expansion of the congruences.  Consequently the equations  for the generalized electric and magnetic fields (generalized vorticity), that are but the appropriate projections of tensor $\mathcal{M}^{\mu\nu}$, turn out to be considerably more involved than previous studies. We have briefly discussed these evolution equations  for a couple of specified geometries. In the process, we derived a relaxed state equilibrium for a plasma coupled to a strongly gravitating source. Gravity gives much more structure to what would have been a London-like state.

The generally covariant formulation provides a basic framework for  investigating the charged fluid dynamics in the presence of strongly gravitating sources when space time curvature might play a significant/dominant role. Our basic equations could be used, for example, to extend the scope and content of numerical simulations of the seed magnetic field generation (\cite{Kulsrud:1996km,Xu:2008yb}).

\section*{Acknowledgment}
The authors are thankful to Justin Feng and David J. Stark for discussion.

\begin{appendix}
	
	\section{Stress Tensor for Perfect Fluid}\label{pfstresstensor}
	The action for perfect fluid is \cite{elze1999variational,brown1993action}
	\begin{equation}\label{actionpf}
	S_{pf}(g_{\mu\nu},n,\tilde{\sigma})=\int d^4x \left(-\sqrt{-g}\rho(n,\tilde{\sigma})\right),
	\end{equation}
	where entropy density $\tilde{\sigma}=ns$ and $n$ and $s$ are the number density and entropy per particle, respectively.
	The variation of the above action is
	\begin{widetext}
		\begin{eqnarray}
		\label{variationaction}
		\delta S=\delta(-\sqrt{-g}\rho(n,\tilde{\sigma}))
		=-\delta(\sqrt{-g})\rho(n,\tilde{\sigma})-\sqrt{-g}\left(\frac{\partial\rho}{\partial n}\delta n+\frac{\partial\rho}{\partial \tilde{\sigma}}\delta \tilde{\sigma} \right)\notag \\ 
		=-\frac{1}{2}\sqrt{-g}\rho(n,\tilde{\sigma})g^{\mu\nu}\delta g_{\mu\nu}-\frac{1}{2}\sqrt{-g}\left(\frac{\partial\rho}{\partial n} n+\frac{\partial\rho}{\partial \tilde{\sigma}}\tilde{\sigma}\right)\left(u^{\mu}u^{\nu}-g^{\mu\nu}\right)\delta g_{\mu\nu}.
		\end{eqnarray}
	\end{widetext}
	Here we have used the conservations laws : $\nabla_{\mu}(nu^{\mu})=0$ and $\nabla_{\mu}(\tilde{\sigma} u^{\mu})=0$ with the four velocity $u^{\mu}=\frac{dx^{\mu}}{ds}$. Now we can write the two following expressions for variations of $n$ and $\tilde{\sigma}$ \cite{Harko:2010zi,LDR}
	\begin{alignat}{2}\label{conservation}
	\delta n=\frac{1}{2}n(u^{\mu}u^{\nu}-g^{\mu\nu})\delta g_{\mu\nu}; \  \delta \tilde{\sigma}=\frac{1}{2}\tilde{\sigma}(u^{\mu}u^{\nu}-g^{\mu\nu})\delta g_{\mu\nu}.
	\end{alignat}
	Therefore, we can simplify equation(\ref{variationaction}) using equation (\ref{conservation}) and get
	\begin{widetext}
		\begin{equation}
		\delta S=-\left[\frac{1}{2}\sqrt{-g}\left(\frac{\partial\rho}{\partial n} n+\frac{\partial\rho}{\partial \tilde{\sigma}}\tilde{\sigma}\right)u^{\mu}u^{\nu}+\frac{1}{2}\sqrt{-g}\left(\frac{\partial\rho}{\partial n} n+\frac{\partial\rho}{\partial \tilde{\sigma}}\tilde{\sigma}-\rho\right)g^{\mu\nu}\right]\delta g_{\mu\nu}.
		\end{equation}
	\end{widetext}
	Now using the definition  $ T^{\mu\nu}=\frac{2}{\sqrt{-g}}\frac{\delta S}{\delta g_{\mu\nu}}$, we obtain the following expression
	\begin{widetext}
		\begin{equation}\label{stress1}
		T^{\mu\nu}=-\left(\frac{\partial\rho}{\partial n} n+\frac{\partial\rho}{\partial \tilde{\sigma}}\tilde{\sigma}\right)u^{\mu}u^{\nu}+\left(\frac{\partial\rho}{\partial n} n+\frac{\partial\rho}{\partial \tilde{\sigma}}\tilde{\sigma}-\rho\right)g^{\mu\nu}.
		\end{equation}
	\end{widetext}
	After Legendre Transformation\cite{marsden1986covariant}, we define the pressure $p$ to be
	\begin{equation}
	p=\frac{\partial\rho}{\partial n} n+\frac{\partial\rho}{\partial \tilde{\sigma}}\tilde{\sigma}-\rho.
	\end{equation}
	This finally gives us the following expressions of stress energy tensor for perfect fluid 
	\begin{eqnarray}\label{stress2}
	T^{\mu\nu}=-(p+\rho)u^{\mu}u^{\nu}+pg^{\mu\nu},\notag \\
	T^{\mu\nu}=(p+\rho)u^{\mu}u^{\nu}+pg^{\mu\nu},
	\end{eqnarray}
	where we have the four velocity $u^{\mu}=\frac{dx^{\mu}}{d\tau}$ with $ds^2=-d\tau^2$ and $c=1$. Alternate expression for the stress tensor can be written by defining enthalpy density $h=p+\rho$ as follows
	\begin{equation}\label{enthalphy}
	T^{\mu\nu}=hu^{\mu}u^{\nu}+pg^{\mu\nu}.
	\end{equation}
\end{appendix}
\bibliographystyle{unsrt}
\bibliography{GR_Feb27}

\end{document}